\documentstyle[12pt,aaspp4]{article}
\received{28 November 2002}
\accepted{24 January 2003}
\slugcomment{Accepted by the Astrophysical Journal Letters} 
\begin{document}

\title{EXPLAINING THE OBSERVED POLARIZATION FROM BROWN DWARFS BY SINGLE DUST
SCATTERING } 
\author{SUJAN SENGUPTA\footnote{sujan@iiap.ernet.in} }
\affil{ Indian Institute of Astrophysics, Koramangala, Bangalore 560 034,
India }

\begin{abstract}
Recent observation of linear optical polarization from brown dwarfs confirms
the dust hypothesis in the atmospheres of brown dwarfs with effective
temperature higher than 1400 K. The observed polarization could arise
due to dust scattering in the rotation induced oblate photosphere or due
to the scattering by non-spherical grains in the spherical atmosphere or
by the anisotropic distribution of dust clouds.
Assuming single scattering by spherical grains in a slightly oblate
photosphere consistent with the projected rotational velocity, the observed
optical linear polarization is modeled by taking grains of different sizes
located at different pressure
height and of different number density. Minimum possible oblateness of
the object due to rotation is considered in order to constrain the grain size.
It is shown that
the observed polarization from the L-dwarfs 2MASSW J0036+1821 and
DENIS-P J0255-4700 can well be explained by several sets of dust parameters
and with the minimum possible oblateness. Models for the observed polarization
constrain the maximum size of grains. It is emphasized that future observation
of polarization at the blue region will further constrain the grain size.

\end{abstract}
\keywords{stars: low mass, brown dwarfs --- polarization --- dust, extinction --- scattering --- stars : atmospheres}


\section{INTRODUCTION}

During the last few years, a large population of L dwarfs covering a range
of effective temperature between 2200 K and 1400 K has been discovered by
several observers ( Martin et al. 1997, Kirpatrick et al 1999). These objects
are characterized by the presence of condensates in their atmosphere.
Theoretical studies (Allard et al 2001, Marley \& Ackerman 2001) reveal that
when the effective temperature of the object is bellow 1300 K, grains are
predominantly formed beyond the visible region of the atmosphere because of
complete gravitational settling. 
At higher effective temperatures however, grains are present in the visible
atmosphere due to incomplete gravitational settling.

Sengupta and Krishan (2001) argued that detectable polarization could
arise due to dust scattering in the atmosphere of L dwarfs where
dust grains should be present in the visible region. In the presence of dust
particulates of small size, the continuum flux in the near optical region
 could be polarized significantly while non-zero polarization can occur in
 the infra-red as well if the particle size is large.
The fast  rotation of the objects (Basri et al 2000) imparts to it the shape
of an oblate ellipsoid and the non-sphericity of the
object should lead to incomplete cancellation of the polarization
of radiation from different areas of the surface.

Very recently detection of linear polarization at 768 nm from a few L dwarfs
has been reported ( Menard, Delfosse \& Monin 2002) that confirms 
dust scattering in the atmosphere of these objects.

In the present letter, I show that single scattering (Mie scattering) by
spherical grains in a slightly oblate photosphere can account the observed
degree of polarization. Observational data for two L dwarfs 2MASSW
J0036+1821 and DENIS-P J0255-4700 is considered from which maximum polarization
is detected.  It is shown that the detected
degree of polarization can well be explained with several  sets of dust
parameters and the minimum possible  oblateness due to rotation. 
An upper limit on the grain size is discussed on the basis of the analysis
of the observed polarization.

\section{ESTIMATION OF THE OBLATENESS DUE TO ROTATION}

 Spectroscopic studies by Basri (1999) and Basri et al. (2000) indicate 
rapid rotation of
brown dwarfs along their axis. A general trend to higher velocities
as one looks to objects of lower luminosity is found in this study.
The projected angular velocity $v\sin i$ for 2MASSW J0036+1821 and
DENIS-P J0255-4700 are determined to be 15.0 kms$^{-1}$ (Schweitzer et al
2001) and $40\pm 10$ kms$^{-1}$ (Basri et al 2000) respectively
 corresponding to an angular velocity of $2.2\times 10^{-4} s^{-1}$ and
$5.8\times 10^{-4} s^{-1}$ respectively. 

The oblateness of a rotating object has been discussed by Chandrasekhar
(1933) in the context of polytropic gas configuration 
and by Hubbard (1984) in the context of solar planets. For
slow rotation, the first order approximation provides the relationship
for the oblateness $q$ of a stable polytropic gas configuration under
hydrostatic equilibrium with mass $M$ and equatorial radius
$r_e$ rotating with an angular velocity $\omega$ as
\begin{eqnarray}\label{obl}
q=\frac{2}{3}C\frac{\omega^2 r_e^3}{GM}
\end{eqnarray}
where $C$ is a constant whose value depends on the polytropic index.
For the polytropic index $n=0$, the density is uniform and the configuration
is known as Maclaurin spheroid. In this case C=1.875. For non-relativistic
completely degenerate gas $n=1.5$ and $C=0.9669$. 

The L dwarfs, DENIS-P J0255-4700 and 2MASSW J0036+1821 belong to the
spectral types L8 (Kirkpatrick et al. 2000) and L3.5 (Reid et al. 2000)
respectively.  Therefore, the effective temperature of these two L dwarfs are
approximately 1420 K and 1870 K respectively (Stephens, Marley, Noll \&
Chanover 2001).
The surface gravity, g, of brown dwarfs varies from $10^5 cms^{-2}$
to $3\times 10^5 cms^{-2}$ (Saumon et al. 1996).
Adopting the empirical relationship given in Marley et al. (1996)
the mass and radius of the two L-dwarfs can be approximated for different
values of g. Hence, by using equation (\ref{obl}), the oblateness $q$ of DENIS-P
J0255-4700 can be estimated as q=0.03 and 0.008
when $g=10^5$ and $3\times10^5$ cms$^{-2}$ respectively
for n=0 (Maclaurin spheroid).
For n=1.5 (non-relativistic completely degenerate gas), q=0.0154 and 0.0042
when $g=10^5$ and $3\times10^5$ cms$^{-2}$
respectively. Similarly, the oblateness of 2MASSW J0036+1821 can be calculated
as q=0.0043 and 0.0012 for n=0 and q=0.0023 and 0.0006 for n=1.5
when $g=10^5$ and $3\times10^5$ cms$^{-2}$ respectively

It is not possible to infer the actual oblateness from
direct observation of brown dwarfs. 
The oblateness decreases
as the polytropic index $n$ increases. Since the density distribution
in brown dwarfs doesn't follow polytropic gas law, the exact estimation
of the oblateness is not possible. Hubbard (1984) showed that $n=1$ is
appropriate for Jupiter.
$n>1.5$ may not be appropriate for brown dwarfs
as it yields much higher pressure.  Further the atmosphere of brown dwarfs
extends even upto 0.005 bar pressure level whereas
in the analysis of Hubbard (1984) the oblateness
at 1 bar pressure level is considered in order to estimate the polytropic
index. Therefore, the values for $q$ with n=1.5 and g=$3\times 10^5$
cms$^{-2}$ can well be considered as
the lowest possible values for the oblateness of the two objects
discussed here. 

\section{POLARIZATION FROM OBLATE PHOTOSPHERE BY SINGLE SCATTERING}

The dependence of polarization due to single scattering by grains on the
oblateness of an object has been discussed by Dolginov, Gnedin \&
Silant'ev (1995) and by Simmons (1982). I follow the formalism given by
Simmons (1982). For optically thin medium the total polarization $p(k)$
can be written as :
\begin{eqnarray}
p(k)=|Z(k)|=|Z^{*}(k)|
\end{eqnarray}
where
\begin{eqnarray}\label{first}
Z(k)=\frac{1}{k^2}\int\int\int{\frac{i_1(\theta,k)-i_2(\theta,k)}{2}n(r,\theta,
\phi)\exp(2i\phi)}d\omega dr.
\end{eqnarray}
Here, $k=2\pi/\lambda$, $\theta$ is the scattering angle and $d\omega$ is the
element of solid angle, $n$ is the particle number density and $*$ denotes
the complex conjugate. $i_1$ and $i_2$
are the scattering functions as defined by van de Hulst (1957). 

If the density distribution function is reasonably smooth then we can write
\begin{eqnarray}\label{second}
n(r,\theta,\phi)=\sum^{\infty}_{l=0}\sum^{m=l}_{m=-l}n_{lm}(r)Y_{lm}(\theta,
\phi)
\end{eqnarray}
where
$P_l^m $ is the associated Legendre function of the first kind and the
function  $Y_{lm}(\theta,\phi)$ is given in Simmons (1982)

Substituting equation (\ref{second}) into equation (\ref{first}) and
integrating over $\phi$ we get
\begin{eqnarray}\label{fourth}
p(k)=\frac{2\pi}{k^2}\sum^{\infty}_{l=2}N_{l2}F_{l2}
\end{eqnarray}
where
\begin{eqnarray}\label{fifth}
F_{lm}=\alpha(l,m)\int^{1}_{-1}\frac{i_1-i_2}{2}P^{m}_{l}(cos\theta)d
(\cos\theta),
\end{eqnarray}
and $\alpha^2(l,m)=(2l+1)(l-m)!/4\pi(l+m)!$.
Considering axisymmetry density distribution with a rotational invariance
around some axis (see Simmons 1982) and using the addition theorem of
spherical harmonic, $N_{lm}$ can be written as 
\begin{eqnarray}
N_{lm}=2\pi\alpha(l,m)P^m_l(\cos i)\exp(-2i\phi)\int^{R_1}_{R_2}n(r)dr
\int^1_{-1}\frac{P_l(\mu)d\mu}{[1+(A^2-1)\mu^2]^{1/2}}
\end{eqnarray}
where $R_1$ and $R_2$ are the outer and inner equatorial axis lengths,
$A$ is the ratio of the length of the equatorial axis to the polar axis
and $\mu=\cos\theta$.

When viewed edge on
$i=\pi/2$ and $\phi=0$ and hence $N_{lm}$ is real. We convert $n(r)dr$
into $n(P)dP$ by using the hydrostatic equation where $P$ is the pressure
at different height of the atmosphere. Therefore,
assuming edge on viewing angle, we can write by using
equation (\ref{fourth}), the degree of polarization as
\begin{equation}
p(k)=\frac{4\pi^2}{k^2 g}\int^{P_2}_{P_1}\frac{n(P)dP}{\rho(P)}
\sum^{\infty}_{l=2}\left[\alpha(l,m)P^m_l(0)F_{l2}(k)\int^1_{-1}
\frac{P_l(\mu)}{\{1+(A^2-1)\mu^2\}^{1/2}}d\mu\right]
\end{equation}
where $g$ is the surface gravity assumed to be constant as
the medium is sufficiently thin and  $F_{lm}(k)$ is given by equation
(\ref{fifth}).
We calculate the integrals in the above equation numerically.
We employed the pressure-density profiles appropriate for the two L-dwarfs
with g=$10^5$ and $3\times 10^5$ cms$^{-2}$. The opacity data is kindly
provided by D. Saumon and the atmospheric opacity sources are discussed
in Saumon et al. (2000). $F_{lm}$  
is also calculated
numerically by first deriving the Mie scattering functions $i_1(\theta)$ and
$i_2(\theta)$ for a given particle size distribution and wavelengths. 
Since  polarization due to small as well as large grains has to be 
investigated, I have included contribution to polarization by multipoles,
l=2,4 and 6.

\section{THE DUST PARAMETERS}

Extensive theoretical studies have been done (Allard et al 2001, Ackerman \&
Marley 2001, Burrows, Hubbard, Lunine \& Liebert 2001) that enlighten
formation, size distribution, density distribution of dust particulates
in the atmosphere of brown dwarfs. However, there is no way to decide 
the most suitable set of grain parameters out of the vast parameter space.
The amount and vertical location of dust should be a function of $T_{eff}$
(Allard et al. 2001) whereas presence of large grains has been predicted
by Ackerman and Marley (2001) and Cooper et al. (2002).

Following  Saumon et al (2000), I have taken a simple 
vertical density profile of the condensate as
\begin{eqnarray}
n_d=AP
\end{eqnarray}
where $n_d$ is the number density of condensed particles, $P$ is the ambient
gas pressure, A is a constant and the cloud layer is bound by
$P_1 < P < P_2$ . The size distribution of particles is given by
\begin{eqnarray}
f(d)=\frac{d}{d_0}\exp\left[\frac{\ln(d/d_0)}{\ln\sigma}\right]^2
\end{eqnarray}
where $d$ is the diameter of the particles and $d_0$ is the mean diameter.
 Dust can form far bellow
the visible atmopshere and there is suggestion that dust may even form
in the convective zone (Helling et al. 2001). Spectroscopic analysis indicates
that beyond about 70-100 bar pressure level the atmosphere becomes invisible.
Therefore, I set the base of the dust cloud at a pressure level 
$P_2=80$ bar. The real part of the refractive index is taken as 1.65 and the
parameter $\sigma$ is fixed at 1.3.
 In order to constrain
the size of the dust particulates I have adopted the minimum possible
oblateness obtained by adopting $g=3\times10^5$ cms$^{-2}$ and the
polytropic index $n=1.5$. The representative sets
of dust parameters that fit the observed polarization from the two objects
are presented in Table~1 and Table~2.

\begin{table*}[htbp]
\caption{Grain parameters for DENIS-P J0255-4700} 
\begin{tabular}{ccccc} \\ \hline
No. & q & $d_0$ ($\mu m$) & A ($cm^{-3}$/bar) & $P_1$ (bar) \\ \hline
1. & 0.0042 & 0.2 & 910.0 & 0.05 \\
2. & 0.0042 & 1.0 & 82.3 & 0.05 \\
3. & 0.0042 & 2.0 & 65.0 & 54.55 \\
4. & 0.0042 & 8.0 & 65.0 & 78.02 \\ \hline
\end{tabular}
\end{table*}

\begin{table*}[htbp]
\caption{Grain parameters for 2MASSW J0036+1821} 
\begin{tabular}{ccccc} \\ \hline
No. & q & $d_0$ ($\mu m$) & A ($cm^{-3}$/bar) & $P_1$ (bar) \\ \hline
1. & 0.0023 & 0.2 & 750.0 & 0.05 \\
2. & 0.0006 & 2.0 & 895.0 & 0.05 \\
3. & 0.0006 & 8.0 & 73.3 & 0.05 \\ 
4. & 0.0006 & 30.0 & 77.0 & 75.0 \\ \hline
\end{tabular}
\end{table*}

\section{RESULTS AND DISCUSSION}
At present there is no clear evidence of magnetic field in L dwarfs. As 
pointed out by Menard et al (2002), it is very much unlikely that the
observed intrinsic linear polarization in the optical arises due to Zeeman
splitting of atomic or molecular lines or due to synchrotron emission. Dust
scattering remains the most probable cause for the polarization from L dwarfs.
There could be more than one way through which the continuum radiation
from L dwarfs would be polarized. The simplest case is scattering by spherical
grains in a rotationally induced oblate photosphere wherein the disc integrated
polarization will not be canceled out. The next possibility is scattering
by non-spherical grains in a perfectly spherical or non-spherical photosphere.
Other possibilities include random distribution of the condensates and
the presence of dust bands.  Since the two L dwarfs have relatively high
rotational velocity, the photospheric disc cannot be perfectly spherical. On
the other hand, a combination of all the possibilities would result into
polarization much higher than that observed. Therefore, I investigate the
simplest possibility, i.e., dust scattering by spherical grains in an
oblate photosphere. This helps to constrain the dust size.
The polarization profiles that fit the observed data are presented in figure~1.


The L dwarf 2MASSW J0036+1821 shows maximum polarization
of 0.199\% among all the candidates observed by Menard et al (2000) 
at 0.768 $\mu m$.  The degree of polarization observed from
DENIS-P J0255-4700 at 0.768 $\mu m$ is 0.167 \%.  
The associated error is 0.028\% and 0.04\% respectively. Now, for the same
rotational velocity, the oblateness increases with the decrease in
surface gravity. As a consequence, degree of polarization is higher for
objects with comparatively lower surface gravity. Further, the integrated
dust number  density is higher for lower surface gravity yielding further
increase in polarization. Therefore, keeping the same dust parameters,
a lower value of polarization can be obtained by increasing the surface
gravity of the object. Hence, with the
same dust parameters, least amount of polarization is produced if the
surface gravity is taken to be $3\times 10^5$ cms$^{-2}$ as the surface
gravity of brown dwarfs varies from $10^5-3\times10^5$ cms$^{-2}$ (Saumon 
et al 1996). This helps to decide the maximum grain size that is allowed
by the observed polarization.
 
Figure~1 shows that the maximum change in the polarization profile occurs
when the mean grain diameter is increased from $0.2 \mu m$ to $ 1 \mu m$.
The degree of polarization peaks near the blue if the grain size is very small
whereas it remains almost the same in the optical region if the grain size
is large. Therefore, if future observation of the same objects detects
higher polarization at blue than that at red then that will imply the
presence of small dust grains. The oblateness of 2MASSW J0031+1821 should
be larger than the minimum possible value in order to yield the observed
polarization by scattering due to small particulates. Obviously, the 
larger is the grain size the less number of particulates should be present.
Thus, degree of polarization serves to constrain the size of grains present
in the atmosphere of brown dwarfs. It is found that if the mean diameter
of grains exceeds about $10 \mu m$ then the observed polarization from
DENIS-P J0255-4700 cannot be
explained by single scattering with spherical dust particulates. However,
2MASSW J0036+1821 can accommodate larger particles as its minimum possible
oblateness is less than that of DENIS-P J0255-4700
owing to low rotational velocity. In reality,
the oblateness could be much higher than its minimum possible value due to
the non-polytropic density distribution and less surface gravity of the
objects. Therefore, the present investigation indicates that the grain
size is unlikely to exceed a few tens of micron. Figure~1 also shows that
the degree of polarization changes significantly with the change in
the values of any parameter e.g., the vertical height or the number density
of grains. Therefore, multiwavelength observation of polarization from the
same object will rule out some of the models providing better idea on
the dust properties and distribution. 

In the present investigation single dust scattering is considered. 
For the earliest type L dwarfs single scattering
approximation is reasonable but multiple scattering could be important for
the late type L dwarfs. Multiple scattering usually lowers the
degree of polarization substantially (Sengupta and Krishan 2001)
because the planes of the scattering
events are randomly oriented and average each other's contribution out
from the emergent polarization. The degree of polarization in
the optical increases slightly if the grain size is increased from 0.1 $\mu m$ 
to 1$\mu m$.  Therefore, in order to obtain the
observed values of the degree of polarization by multiple dust scattering,
not only very large grains are needed but the number density of grains must
also be very high. Since the number density of grains should not exceed the
mass of the heavy elements, multiple scattering may not give rise to sufficient
amount of polarization unless the oblateness is high. However, the effect
of multiple scattering with large grains is worth investigating as it will
provide information on the contribution of molecular absorption as well to the
polarization profile.

\section{CONCLUSION}

Observation of polarization and its theoretical analysis serve as a potential
tool to understand the properties of condensates in the atmosphere of brown
dwarfs. In addition to that it provides important information on the
geometrical asymmetry of the photosphere. The important message that is
conveyed in this letter is that the observation of polarized radiation can
very well constrain the grain size. Modeling of the observed polarization from
two L dwarfs indicates that the mean diameter of grains is unlikely to
exceed a few tens of microns.
Also, the surface gravity of the objects plays
a crucial role in determining the degree of polarization. Lower surface gravity
yields higher asymmetry in the photospheric disk as well as higher number
of dust particulates. Both help in increasing the degree of polarization
significantly. Therefore, if the surface gravity of brown dwarfs is less,
large grains cannot be accommodated in the atmosphere.
The oblateness of the object is 
constrained by the observed rotational velocity. Single dust scattering
is consistent with the available knowledge on the oblateness and grain size of
brown dwarfs. Since multiple scattering 
yields much less polarization, larger grain size with greater number density
and greater oblateness are
needed to account the observed polarization. Observation of variable 
polarization should indicate randomly distributed dust cloud.  
Further observation of
polarization at different wavelengths will provide significant information
on the grain properties in the atmosphere of brown dwarfs.

\acknowledgments
I am grateful to the anonymous referee for several useful comments, suggestions
and constructive criticisms that have improved the quality of the paper in
great extent.
 I am thankful to Francois Menard for some discussion on the observed data
and to D. Saumon for providing the brown dwarf opacity data.
 Thanks are due to M. Parthasarathy and  A. V. Raveendran, for useful
suggestions and discussion and to
V. Krishan and  N. K. Rao for their encouragement.

\clearpage

\plotone{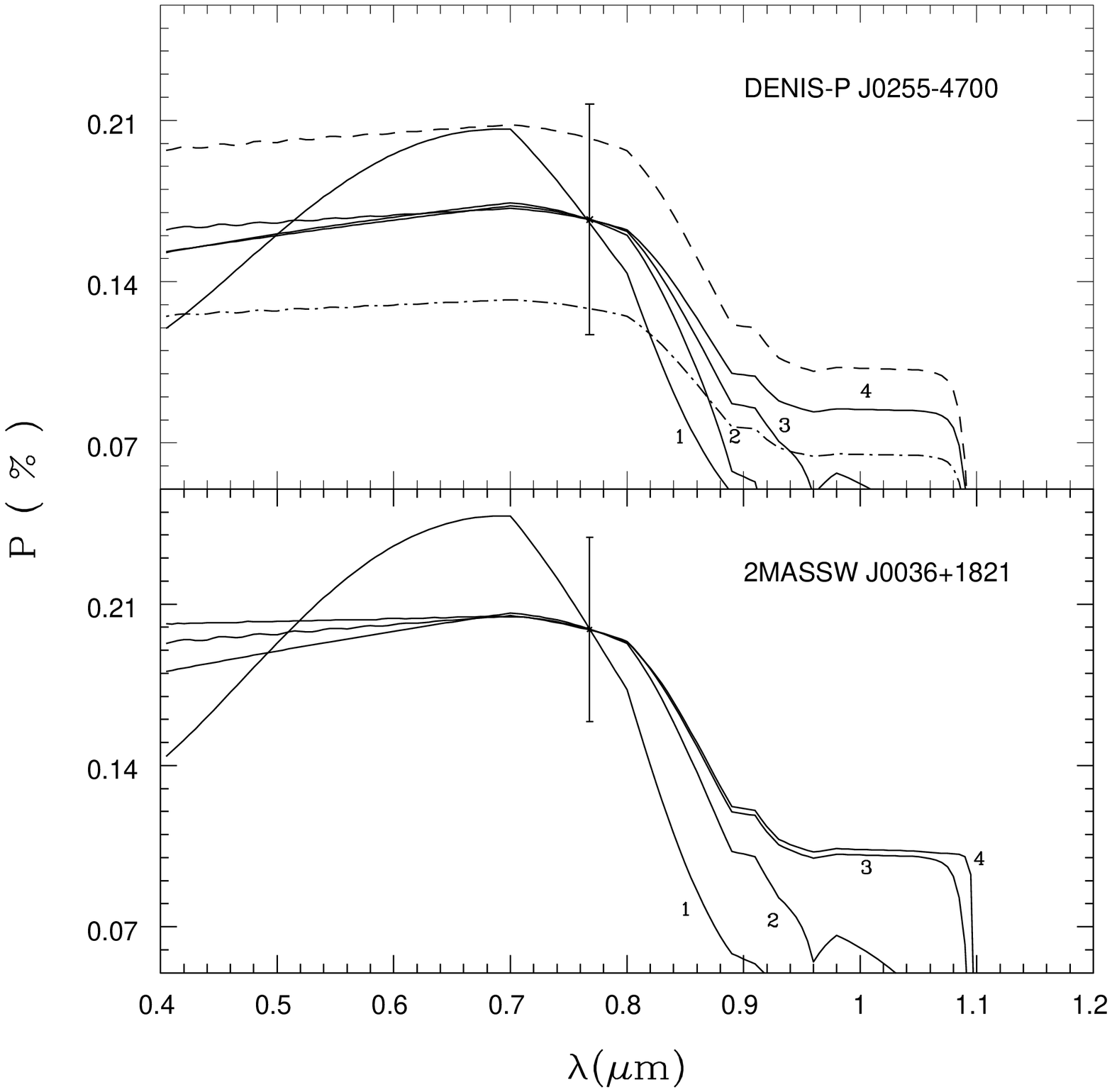}
\figcaption[f1.eps]{Degree of polarization as a function of wavelength for
the two L dwarfs. The numbers associated with each curve represent the set
of grain parameters given in Table~1 and Table~2. The dashed line
represents model 4 for DENIS-P J0255-4700 but with $P_1=77.6$ bar.
The dot dashed line represents model 4 for the same object but with
$A=50.0$ cm$^{-3}$/bar.  The observed polarization is at 0.768 $\mu m$.
\label{figure1}}

\end{document}